\documentclass{pasa}%

\title[Correlations between kinematics and metallicity]{Correlations between kinematics and metallicity \\in the Galactic bulge: a review}
\author[Carine Babusiaux]{Carine Babusiaux$^1$\\
\affil{$^1$GEPI, Observatoire de Paris, PSL Research University, CNRS, Universit\'e Paris Diderot, Sorbonne Paris Cit\'e; 5 Place Jules Janssen 92195 Meudon, France}}%
\jid{PASA}
\doi{10.1017/pas.\the\year.xxx}
\jyear{\the\year}

\usepackage[authoryear]{natbib}
\usepackage{aas_macros}
\usepackage{xcolor}

\bibpunct{(}{)}{;}{a}{}{,}
\def\degr{^\circ}
\newcommand{\kms}{{\mathrm{km~s^{-1}}}}
\newcommand{\FeH}{{\mathrm{[Fe/H]}}}

\newcommand{\alphaFe}{{[\alpha/\mathrm{Fe]}}}
\newcommand{\Vr}{{V_\mathrm{R}}}

\begin{document}%
\begin{abstract}
Recent large scale surveys of galactic bulge stars allowed to build a detailed map of the bulge kinematics. The bulge exhibits cylindrical rotation consistent with a disky origin which evolved through bar driven secular evolution. 
However correlations between metallicity and kinematics complicate this picture. In particular a metal-poor component with distinct kinematic signatures has been detected.
Its origin, density profile and link with the other Milky Way stellar populations are currently still poorly constrained. 
\end{abstract}
\maketitle%
\section{Introduction}

Two main scenarios, with very different signatures, have been invoked for bulge formation. The first scenario
corresponds to the gravitational collapse of a primordial gas and/or to the hierarchical merging of sub-clumps and leads to what is called a classical bulge. 
In those cases the bulge formed before the disc and the star formation time-scale was short. The resulting
stars are old, present enhancements of $\alpha$-elements relative to iron and isotropic kinematics. 
The second scenario is secular evolution of the disc through a bar forming a pseudo-bulge. The bar heats the disc in the
vertical direction, giving rise to the typical boxy/peanut aspect. The resulting bulge presents bar driven kinematics and age and chemistry corresponding to the properties of the disc at the bar formation time. Further star formation are expected to be induced by the bar driving gas towards the galactic centre and by shocks at the bar end. 

The first suggestion for the presence of a bar in the Galactic inner regions came from gas kinematics \citep{deVaucouleurs64} showing strong non circular motion. 
The presence of a bar is now firmly confirmed both from the overall peanut shape of the bulge and its kinematics. %The chemistry and the ages studies indicates an old $\alpha$-enhanced population for the bulge. 
But is there also a more primordial population in the bulge which could correspond to the first scenario? 
Recent large scale studies of the bulge kinematics are trying to answer this question.
We will review them here together with some studies of specific tracers. Section \ref{sec:young} and \ref{sec:old} review the kinematics of the youngest and oldest bulge populations respectively. In section \ref{sec:global} the global kinematics of the bulge through large surveys is summarized. Section \ref{sec:stream} presents the detection of bar induced streaming motion. Section \ref{sec:metvar} study the global variation of kinematics as a function of metallicity for the main populations. Section \ref{sec:pops} study the different interpretations in terms of population of those different kinematic behaviour. A short conclusion is given in section \ref{sec:conclu}.

\section{\label{sec:young}Kinematics of young metal-rich population tracers}

\begin{figure}[t]
 \centering
  \includegraphics[width=8cm]{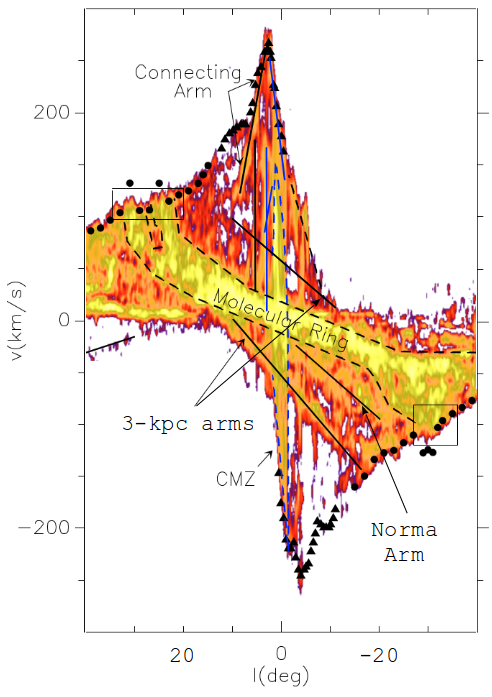}
 \caption{Figure adapted from \cite{Nemesio08}: Longitude-velocity (lv) diagram of the CO(1-0) emission \citep{Dame01}. The solid lines trace the position of some remarkable features
such as the locus of the spiral arms, the 3-kpc near and far arms and the Connecting Arm. The black dashed lines indicate the contour of the Galactic Molecular
Ring. The solid circles and triangles are the terminal velocities measurements. The boxes mark the position of the spiral arms tangent points. The lines concerning the Nuclear Disk or Central Molecular Zone are in blue.}
 \label{fig:Nemesio}
 \end{figure}

% The gaz
The youngest stars lie close to their birth place and follow the main gas structures. The gas kinematics are represented in the well known longitude-velocity diagram illustrated in Fig. \ref{fig:Nemesio}. 
The Galactic Molecular Ring (GMR) dominates the gas longitude-velocity diagram. This prominent feature marks a region
of enhanced molecular density at Galactocentric radii between 4 and 6~kpc and is usually used to define the ``inner'' regions of our Galaxy. It may not be an actual ring but the inner parts of the spiral arms \citep[e.g.][]{Dobbs12}. 
The near and far 3-kpc arms are symmetric lateral arms that contour the bar \citep{Dame08}.
The Central Molecular Zone (CMZ) seems to be the gas response to an inner bar \citep{Sawada04,Nemesio08} and could correspond to the inner Lindblad resonance of the main bar. 
While the main bar is supported by x1 orbits (elliptical orbits which are more elongated towards the inner regions, see Fig. \ref{fig:barschema}), the CMZ seem to mark the switch to the x2 orbit family (orbits lying inside the inner most cusped x1 orbit, oriented perpendicular to the x1 orbits). 
%While the main bar is supported by x1 orbits, the CMZ seem to mark the switch to the x2 orbit family. 
The Connecting Arm could correspond to the bar near-side dust lane created by the shock front and plunging towards the CMZ \citep{Fux99}. 

% Maser stars
% OH/IR ABG star / maser sources in late type stars and star forming regions 
Maser emissions, seen around star forming regions and young AGB stars (with ages typically 0.2 to 2 Gyr), are tracing young galactic structures.
%Maser stars are relatively young  and metal-rich AGB stars. 
Masers distribution in the lv diagram, for both OH and SiO masers, shows an overall agreement with the gas one and the presence of forbidden regions, in agreement with a motion in a bar like potential \citep[e.g.][]{Deguchi04,Habing06}. 
%Note that with OH maser samples containing more star forming regions and therefore following more the gaz strong features than SiO stars .
Masers are also observed at relatively high latitude ($\vert b \vert>5 \degr$).
\cite{Izumiura95} noted the difference in radial velocity dispersion with latitude, the higher dispersion being in the inner regions. 
Parallaxes to some of those masers stars are now available allowing a 3D positioning of the lv diagram features \citep{Sanna14}. Both their velocities and their spatial distribution indicate that those stars belong to the Galactic bar and they have been largely used to derive bar parameters \citep[e.g.][]{Sevenster99, Debattista02}.

% Classical Cepheids 
Classical Cepheids, standard candles with ages lower than $\sim300$~Myr would be ideal tracers of the kinematics of the youngest stars. Currently only nuclear disk Cepheids, where they are concentrated, have been kinematically studied and compared with the gas kinematics \citep{Matsunaga15}.

\section{\label{sec:old}Kinematics of old metal-poor population tracers}

The full bulge seems to be mainly old \citep{Kuijken02,Zoccali03,Clarkson08,Brown10,Valenti13}. We therefore only present in this section the most classical tracers of the oldest and the more metal-poor populations.

The globular cluster metallicity distribution has been found to be doubled peaked at mean $\FeH \simeq -1.6$ and $\FeH \simeq -0.5$. The inner metal-rich population shows rotation, with kinematic signatures comparable to field stars and may be associated with the bulge \citep[e.g.][]{Cote99,Dinescu03,Rossi15} while the metal poor clusters have kinematics more consistent with a halo component.

The bulge RR Lyrae population is centred around a metallicity of $\FeH=-1$~dex \citep{WalkerTerndrup91,Pietrukowicz12} and has a much more spheroidal, centrally concentrated distribution then the bulge red clump stars, but seems to follow the boxy-peanut shape in the inner regions \citep{Pietrukowicz12,Dekany13}. Kinematic studies of those RR Lyrae stars are not yet available apart from those of Baade's Window showing a velocity dispersion of 133$\pm25$~$\kms$ \citep{Gratton87}. 
Similarly Type II Cepheids are centrally concentrated \citep{Soszynski11Cepheids} and the distribution of old short-period Miras appears spheroidal in the outer bulge, opposite to the young long-period ones \citep{Catchpole16} but no kinematic studies of those are available yet. 

Field K giant studies of \cite{Minniti96} and \cite{Ness13kine} found that the most metal-poor stars, with $\FeH<-1.0$, presents no significant or small rotation (respectively) and a velocity dispersion around $120~\kms$ independent of Galactocentric distance. Such metal-poor stars have been very recently found also in the galactic inner regions \citep{Schultheis15,Do15}. We will discuss further the metal-poor tail of the field stars in section \ref{sec:metvar}.

\section{Global kinematics of field stars\label{sec:global}}

% Vr without metallicity: 
Some other kinematic studies of specific bulge stars such as Mira variables \citep{Menzies90}, carbon stars \citep{TysonRich91} or planetary nebulae \citep{Durand98} found rotation and a decrease of velocity dispersion with increasing distance from the galactic plane, but the underlying age and metallicity of those stars are difficult to assess. 
The population of bulge Planetary Nebulae include young stars \citep{Buell13} and seem to poorly represent the metal-poor population \citep{Gesicki14}. Their kinematics have been studied by \cite{Beaulieu00} to derive the galactic bar properties. However \cite{Uttenthaler12} note that their radial velocities at high galactic latitude (b=-10$\degr$) behaves like the metal-poor population. 

M and K-giant bulge stars are now the targets of large scale kinematic surveys. The first one is the Bulge Radial Velocity Assay survey (BRAVA; \citealt{Rich07brava, Howard09, Kunder12}), providing radial velocities of 9,500 2MASS M giants at  $\vert l \vert < 10\degr$ and b = -4, -6, -8$\degr$. The Abundances and Radial velocity Galactic Origins Survey (ARGOS; \citealt{Freeman13,Ness13met,Ness13kine}) measured radial velocities, $\FeH$ and $\alphaFe$ ratios for $\sim$28,000 stars, targetting 2MASS K-giants and observing mainly at $\vert l \vert < 20\degr$ and b=-5, -7.5, -10$\degr$. 
The Giraffe Inner Bulge Survey (GIBS; \citealt{Zoccali14}) is a survey of $\sim$6500 VISTA VVV red clump giants within $\vert l \vert \leq 8\degr$ and $\vert b \vert \leq 8\degr$; radial velocities have been published \citep{Zoccali14} as well as $\FeH$ and $\alphaFe$ ratios for the high-resolution fields at b=-4$\degr$ \citep{Gonzalez15} and it will soon be completed by metallicities and in the near future by VVV proper motions. 
Preliminary results have also been published on larger scale spectroscopic surveys that include the bulge in their field selection: the APOGEE survey (\citealt{Majewski15}, targetting 2MASS M-giants: \citealt{Nidever12}) and the Gaia-ESO survey (GES; \citealt{Gilmore12}, targetting VISTA K-giants: \citealt{Rojas14}). 

The BRAVA data showed that the bulge stars follow a cylindrical rotation, e.g. their mean rotation speed is roughly independent of the high above the disc. Using different target selections, the ARGOS and GIBS surveys results are consistent with the BRAVA ones. Their radial velocity distribution is very well reproduced by a N-body model of a pure-disc Galaxy \citep{Shen10}. 

% proper motion without metallicity:
Stellar proper motion studies have also been made in small fields either on RGB stars \citep{Spaenhauer92, Vieira07, Vasquez13} or on main-sequence stars using HST \citep{Kuijken02,Kozlowski06,Clarkson08,Soto14}, as well as on a large survey area through the OGLE proper motions \citep{Sumi04,Rattenbury07,Poleski13}. Those studies shows a velocity anisotropy consistent with bar models \citep[e.g.][]{Zhao96,Qin15}.

\section{The streaming motion\label{sec:stream}}

\begin{figure}[t]
 \centering
  \includegraphics[width=6cm]{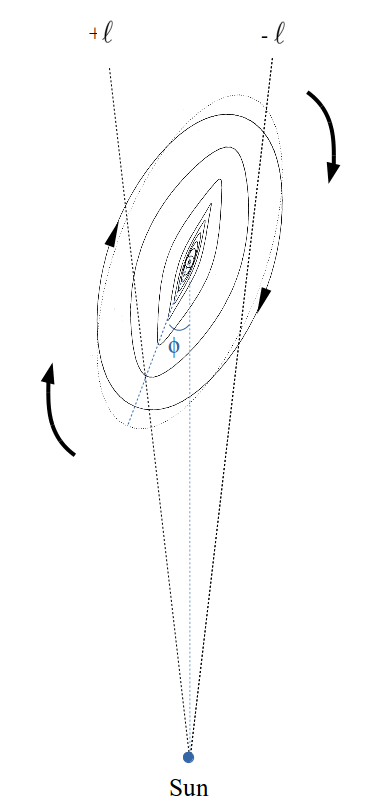}
 \caption{Schematic representation of the main bar driven motions. The schema is adapted from \cite{Athanassoula92} showing some periodic orbits of the x1 family which are the backbones of the bar together with the bar outline as a dotted line. The schema has been oriented at a bar angle $\phi$ of 20$\degr$.}
 \label{fig:barschema}
\end{figure}

A difference of the velocities as a function of distance is expected in a bar potential. Figure \ref{fig:barschema} illustrates the expected streaming motion due to the x1 family orbits in the plane. As the stars form, the bar streams in the same sense as the Galactic rotation, and because the bar is in the first quadrant, the stars on the near side of the bar are expected to go towards us, while stars on the far side should move away from us. Note that the actual velocity shifts between these two streams constrains the bar orientation angle \citep{MaoPaczynski02}.

\begin{figure*}[t]
 \centering
  \includegraphics[width=12cm]{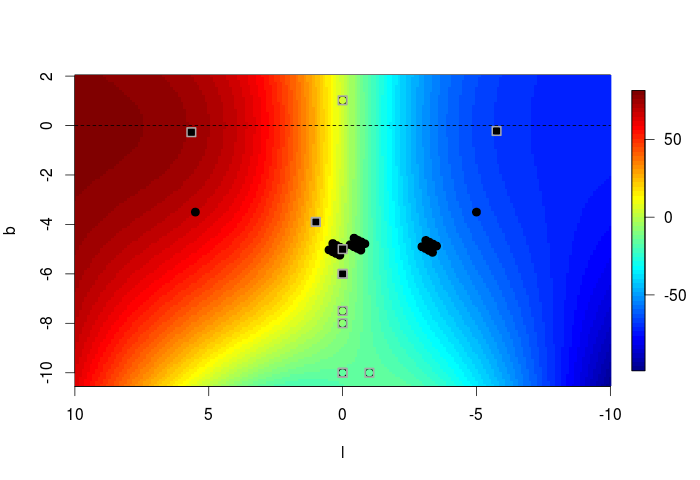}
 \caption{Detection of streaming motion. Background: Mean radial velocity surface in the longitude-latitude plane as derived from the GIBS Survey \citep{Zoccali14}. Filled symbols: streaming motion detected. Open symbols: streaming motion un-detected. Grey box: fields where metallicity information was used. Compilation from \cite{Rangwala09a,Babusiaux10,DePropris11,Uttenthaler12,Ness12,Vasquez13,Poleski13,Babusiaux14,Rojas14}}.
 \label{fig:streamingmotion}
 \end{figure*}
 
In the absence of direct distance information, streaming motions have been studied through
velocity variation as a function of magnitude \citep[e.g.][]{Rangwala09a}, specifically targetting the bright and the faint red clump peaks corresponding to the X-shape when they are separately detected \citep[e.g.][]{Vasquez13} or using statistical distances \citep{Babusiaux14}.
At low latitude ($\vert b \vert \leqslant 6 \degr$) a difference in radial velocity ($\Vr$) and/or proper-motion as a function of distance has been seen by those techniques \citep{Rangwala09a,Babusiaux10,Ness12,Vasquez13,Poleski13,Babusiaux14,Rojas14}. 
However at higher latitude, no strong difference in the kinematics for the two red clumps have been found \citep{DePropris11,Uttenthaler12,Ness12,Rojas14}. This difference between low and high latitude is predicted by bar models \citep{Ness12}.
The location of those studies of the streaming motion are overlaid on the overall map of mean radial velocity obtained from the GIBS survey in Fig. \ref{fig:streamingmotion}. The fields studied with chemistry information are highlighted by grey box. Note that \cite{Rangwala09a}, observing at $b\sim3.5\degr$ saw $\Vr$ variations with magnitude at $\vert l \vert=\pm$5$\degr$ but not in Baade's Window (b=1$\degr$); \cite{Babusiaux10} and \cite{Rojas14} also did not find difference in $\Vr$ in BW for the full sample but a difference started to be significant when removing the metal-poor component of the sample. No streaming motion is detected at (l,b)=(0$\degr$,1$\degr$); the kinematics there may be dominated by the nuclear bulge \citep{Babusiaux14}.

 The high-velocity peaks detected by \cite{Nidever12} at $\Vr \sim 200~\kms$ are likely associated to the streaming motion too. High-velocity components are also seen in \cite{Babusiaux14} fields at (l,b)=($\pm6\degr$,0$\degr$), presenting a mean metallicity of 0.2~dex (e.g. corresponding to population A in \citealt{Ness13kine}, see section \ref{sec:metvar}). The velocity versus distance trend indicates that the high-velocity components are behind the main bar component at l=+6$\degr$ and in front of it at l=-6$\degr$, confirming the \cite{Nidever12} interpretation of this high-velocity component as being linked to the bar dynamics (see also \citealt{AumerSchonrich15} and references therein). Such high velocity peaks are not seen in the GIBS survey \citep{Zoccali14} although observing also in low latitude fields (reaching b=$-2\degr$). This could be explained by the strong difference in target selection. In particular the APOGEE survey presents a much larger contamination by foreground disc stars, which counter-intuitively leads to a higher highlight of the high velocity component: the foreground disc stars have a mean radial velocity peak around zero that do not spread out to $\Vr \sim 200~\kms$, the large velocity tail is therefore picked-up as a separate component by a Gaussian decomposition, while the GIBS radial velocity profiles have a large dispersion out to the high velocity peak, consistent with one single Gaussian with a large dispersion. 

When detected, the streaming motion is associated with the metal-rich component. The only exception being \cite{Vasquez13} observing at (l,b)=(0$\degr$,-6$\degr$)  who found that the velocity difference between the bright and the faint clumps is larger for stars with sub-solar metallicity values. This field can be seen in Fig. \ref{fig:streamingmotion} as being the lowest latitude field where the streaming motion is detected. In this field the metal-rich component starts to be less prominent and may suffer more from foreground stars contamination. \cite{Rojas14} observing the same field confirmed the association of the streaming motion to the metal-rich component. \cite{Rojas14} has a broader target selection in colour allowing to probe the metal-poor tail, while the \cite{Vasquez13} MDF is actually dominated by the component called B in \cite{Ness13kine} study with a mean metallicity of -0.25 that they associate with the thick boxy/peanut-bulge. 

In fact, for high latitude studies, the double clump is itself biased towards metal-rich stars as the double clump is not seen when selecting only metal-poor stars ($\FeH<-0.5$) \citep{Ness12,Uttenthaler12,Rojas14}. This could be due to a real structure difference between metal-rich and metal-poor stars and/or stellar evolution bias on the red clump morphology \citep{Nataf14}.

\section{Kinematics as a function of metallicity for bulge giants\label{sec:metvar}}
% Vr dispersion With metallicity
%The velocity dispersion change with metallicity
Early observations of bulge giants \citep{Rich90,Sharples90,Spaenhauer92,Minniti96} already found that the most metal-rich stars have a smaller velocity dispersion, all observing at $\vert b \vert \geqslant 4\degr$. 
More metal-rich stars have also a larger apparent anisotropy $\sigma_l / \sigma_b$, a smaller $\sigma_b$, and $\sigma_r > \sigma_b$ when compared to metal-poor stars in Baade's Window \citep{Soto07, Babusiaux10}.
\cite{Vieira07} observing at (l=0$\degr$,b=-8$\degr$) found that blue horizontal branch stars (tracing an old metal-poor population) showed a higher dispersion in both l and b but the same anisotropy $\sigma_l/\sigma_b$ as the main RGB sample.

\begin{figure}[t]
 \centering
  \includegraphics[width=9cm]{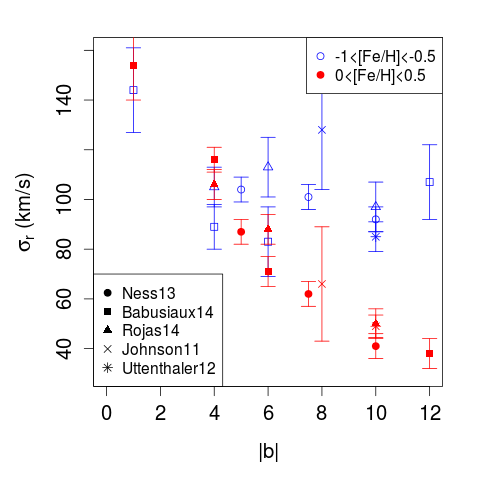}
 \caption{Radial velocity dispersion as a function of latitude along the bulge minor axis ($\vert l \vert \leqslant 1\degr$) for metal-rich stars ($0<\FeH<0.5$, blue open symbols) and metal-poor stars ($-1<\FeH<-0.5$, red filled symbols) as derived from:{\it circle symbols:} the ARGOS survey by \cite{Ness13kine}; {\it squares:} \cite{Babusiaux10} and \cite{Babusiaux14}, as already compiled in the later; {\it triangles:} the GES iDR1 \citep[][removing the \cite{Hill11} targets used in the figures of this paper]{Rojas14}; {\it cross:} \cite{Johnson11}; {\it star:} \cite{Uttenthaler12}.}
 \label{fig:sigVrlat}
 \end{figure}

Along the bulge minor axis, it has been shown that the metal poor population presents the same velocity dispersion, while the metal rich population goes from bar-like high velocity dispersion to disc-like low velocity dispersion while moving away from the galactic plane.
A compilation is presented in Figure \ref{fig:sigVrlat} from several different surveys with different target selection and analysis, pointing towards a quite robust result.  
Baade's Window seems to be at the transition where the metal-rich stars start to have a higher velocity dispersion than the metal-poor ones closer to the plane. A small difference in target selection bias could therefore lead to the inverted trend in the results of \cite{Rich90} and \cite{Babusiaux10} and the flat velocity dispersion profile of \cite{Soto07}, all observing in Baade's Window. The increase of velocity dispersion closer to the galactic plane along the minor axis is predicted by bar dynamical models and confirmed observationally by the GIBS survey \citep{Zoccali14}. At high latitude, the metal-poor population is the dominant one which explains the high velocity dispersion of AGB stars and planetary nebulae noted in \cite{Uttenthaler12} in the field at b=-10$\degr$. % see their fig. 20.

Note that the inner bulge M-giant sample of \cite{Rich12} and the AGB sample of \cite{Uttenthaler15} could not be added to Fig. \ref{fig:sigVrlat} as the metal rich and the metal poor tails of the bulge metallicity distribution function are not present in their  data. It has been suggested that very high mass-loss rates at high metallicities could remove those stars from the canonical paths of stellar evolution \citep[e.g.][]{Cohen08}, % ; Castellani & Castellani 1993), 
hence they do not reach the coolest and most advanced phases (see discussion in \citealt{Uttenthaler15}). A more detailed simulation of the biases between those different target selections would be important to understand the differences between different studies.

The strong decrease in velocity dispersion with increasing metallicity at high latitude is not only seen along the minor axis (Fig. \ref{fig:sigVrlat}) but also in off-axis fields \citep{Ness13kine,Johnson13}.

The metal-rich velocity dispersion change is in agreement with the SiO masers measurements of \cite{Izumiura95} indicating a velocity dispersion of 109~$\kms$ for $3 < \vert b \vert < 5\degr$ and 68~$\kms$ further away from the plane. 

The large scale ARGOS survey \citep{Ness13kine} dissect the bulge metallicity distribution in three main components: A is the most metal-rich with $\FeH \sim +0.15$ that they associate with a relatively thin and centrally concentrated part of the boxy/peanut bulge. B with a mean $\FeH \sim -0.25$ is a thicker boxy/peanut bulge with a relatively constant fraction within $\vert l \vert < 15\degr$. A seems to be a colder more compact version of B. C with a mean $\FeH \sim -0.7$ still shows significant rotation but has an overall different behaviour in its kinematics versus A and B with a latitude independent dispersion within $\vert l \vert < 10\degr$. At $\FeH<-1$ the rotation is lower than for stars with $\FeH>-1$ by about 50 percent. 

\begin{figure}[t]
 \centering
  \includegraphics[width=6cm]{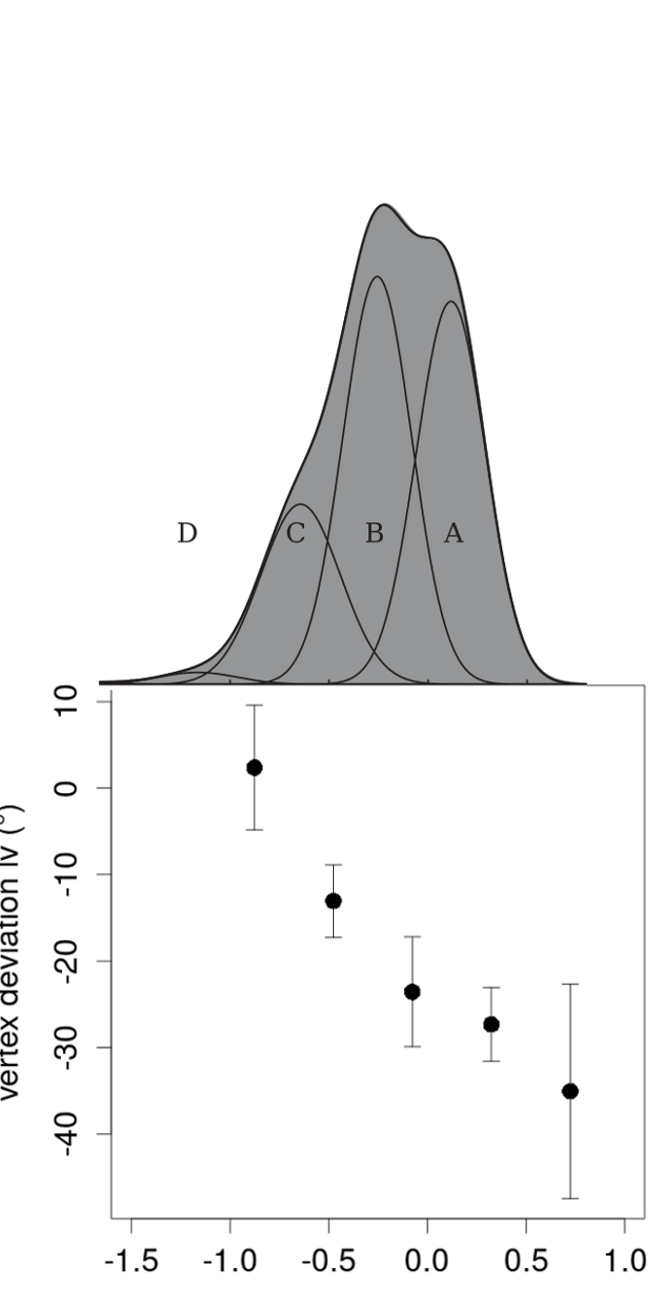}
 \caption{Top: ARGOS MDF decomposition at b=-5$\degr$ (From \cite{Ness13met}). Bottom: Vertex deviation in Baade's Window (l=1$\degr$,b=-4$\degr$) as a function of metallicity by bins of 0.4 dex, data compilation from \cite{Babusiaux10} and GES iDR1 \citep{Rojas14} : 542 stars from three different target selections \citep{Zoccali08,Hill11,Rojas14} with OGLE-II proper motions \citep{Sumi04}. } 
 \label{fig:vertexdev}
 \end{figure}

% Vertex deviation
The vertex deviation, e.g. correlation between the radial and the longitudial velocities measured by the orientation of the axis of the velocity dispersion ellipsoid in the radial-longitudinal velocity plane, is characteristic of a nonaxisymmetric system. 
\cite{Zhao94}, \cite{Soto07} and \cite{Babusiaux10}, observing red giant stars in Baade's Window, found evidence of vertex deviation in the most metal-rich stars ($\FeH \gtrsim -0.5$). 
Figure  \ref{fig:vertexdev} presents an update of the figure of \cite{Babusiaux10} done by adding the GES iDR1 data \citep{Rojas14}. 
The exact vertex deviation angle is dependant on the relative number of stars before or after the distance assumed in the proper motion derivation. The exact location of the break in metallicity is also difficult to quantify as only 30\% of star showing a vertex deviation can introduce a significant correlation in the full sample. In the same way, the mean rotation and the velocity dispersion of a sample depends on its contaminants. The top of Fig. \ref{fig:vertexdev} indicates the location of the main bulge populations found by \citep{Ness13met}. Both studies may not be exactly on the same metallicity scale but this representation can help the interpretation in term of population (see next section) as those population mix in given metallicity ranges. For example, global rotation and vertex deviation in the metallicity range $-1<\FeH<-0.5$, range typically associated with population C, should be higher than the real one of population C due to population B contamination. Detailed comparisons with models should take this into account together with sample selection bias.

\section{\label{sec:pops}Interpretations in terms of populations} 

The different kinematics as a function of metallicity indicates a composite nature of the bulge. The metal-rich population of the bulge follows bar-driven kinematics. The metal-poor population has a distinct kinematic behaviour, presenting a smaller rotation, a roughly constant velocity dispersion and no vertex deviation. Other constraints on the shape, abundances and ages of those populations are also available, which are necessary for the interpretation of those differences in terms of galactic populations (see the other reviews in this special issue for more details). 

% shape
The metal-rich population is more concentrated closer to the galactic plane and clearly follows the X-shaped boxy bulge. 
The presence of a second, thinner bar complexify the structure of the metal-rich population although this long bar could simply correspond to
the leading ends of the bar in interaction with the adjacent spiral arm heads \citep{MartinezValpuestaGerhard11,RomeroGomez11}.
The metal-poor population is centrally concentrated, extends further from the plane and does not follow (as tightly as least) the X-shape structure (see references in section \ref{sec:old} for the oldest tracers and section \ref{sec:stream} for the discussion on the un-detected double peak in the metal-poor giant branch). 
The changing ratio of these populations with latitude would lead to an apparent vertical metallicity gradient \citep{Zoccali08,Ness13met}. However an initial radial metallicity gradient could also lead to an observed vertical gradient through secular evolution alone \citep{MartinezValpuestaGerhard13}. The metallicity gradient is visible at $\vert b \vert > 4\degr$ but flattens in the inner regions \citep{Ramirez00,Rich07inner}. The metal-poor component has a significant contribution at all latitudes including in the inner regions where it seems mixed with the metal-rich one \citep{Babusiaux14}.

% abondances
The metal-poor population shows high [$\alpha$/Fe] ratios while the metal-rich population shows a low $\alphaFe$ roughly similar to the thin disc \citep[e.g.][]{Hill11}. This could be explained with a metal-rich population formed on a longer time-scale and with lower star formation efficiency, consistent with the secular formation scenario, while the metal-poor population formed on a very short time-scale by means of an intense burst of star formation of high efficiency \citep[e.g.][]{Grieco12}. 
Chemical similarities between the metal-poor part of the bulge and the thick disc have been highlighted \citep{Melendez08, AlvesBrito10,Bensby13,Gonzalez15}.

%ages
From microlensed bulge dwarfs a difference in the age distribution is also seen: the metal-rich stars show a wide distribution in ages while the metal-poor stars are all old \citep{Bensby13}. 
Although the bulge is mainly old, an extended star formation history is required to explain the more massive AGB stars in the bulge \citep{vanLoon03,GroenewegenBlommaert05,Uttenthaler07}. 
\cite{ColeWeinberg02} indicate an upper age of 6~Gyr for the bar from its infrared carbon star population but those may instead be mass-losing O-rich stars (see e.g. the discussion in \citealt{Catchpole16}).
While the young stars of the bulge can be easily associated with a bar driven secular evolution of the disc, the situation of the old stars is not that clear. A mix of old and young stars is indeed expected in a secular evolution scenario, with the youngest stars being closer to the plane \citep[e.g.][]{Ness14}. 
\\
% the models

Several models study the hypothesis of a double composition of the bulge in detail \citep[e.g.][]{Samland03, Nakasato03, Athanassoula05, Rahimi10, BekkiTsujimoto11, Grieco12, Robin12, Perez13}, where the metal-rich population is associated with the bar and the metal-poor population is either the thick disc or a primordial structure formed either by hierarchical formation, dissipational collapse, or clumpy primordial formation. 
In particular the formation time, strength and longevity of the bar have been shown to be dependant on the properties of the galaxy at high redshift \citep[e.g.][]{Kraljic12,Athanassoula13}.

\cite{Shen10} reproduced the BRAVA radial velocity distribution with a N-body model of a pure-disc Galaxy, concluding that any classical bulge contribution
cannot be larger than $\sim8\%$ of the disc mass. Comparing their model with the ARGOS survey, \cite{DiMatteo14} also exclude a massive classical bulge and give some insight on the kinematic and proportion difference between populations A and B, indicating that population A may have formed on average closer to the Galaxy centre than the component B.  

Other dynamical models have shown that the formation of a bar could spin-up to a faster rotation a small classical bulge already present before the bar formation \citep{Saha12}. It would be very difficult a posteriori to detect its presence via kinematics alone and indeed chemistry/kinematics correlations could highlight this component. % \citep{Gardner13}
The hint for the elongation of the inner RR Lyrae stars and the rotation of the inner metal-rich globular clusters could be consistent with this angular momentum transfer between the bar and an initial classical bulge during their co-evolution.

The model of \cite{Fux99} reproduces all the main trends of the bulge shape and kinematics, including the gas one, but opposite to the model of \cite{Shen10} it has a massive spheroid component (0.5 times the mass of the disc). The proportion of spheroid versus disc stars is roughly consistent with the observations in Baade's Window \citep{Babusiaux10}. The mean rotation of the spheroid is lower than the disc one by about 50 per cent, similar to what is observed for the metal-poor stars in \cite{Ness13kine}. There was not as much radial velocity measurements available at the time to constrain this model and it over-predicts the velocity dispersion by $\sim 14$\% while the most massive classical bulge of \cite{Shen10} (0.3 times the mass of the disc) over-predicts the velocity dispersion by already 23\%. The number of parameters for such models are large and it is quite logical that a model that does fit the data without a classical bulge does not anymore when a massive classical bulge is added.

\begin{figure}[t]
 \centering
  \includegraphics[width=6cm]{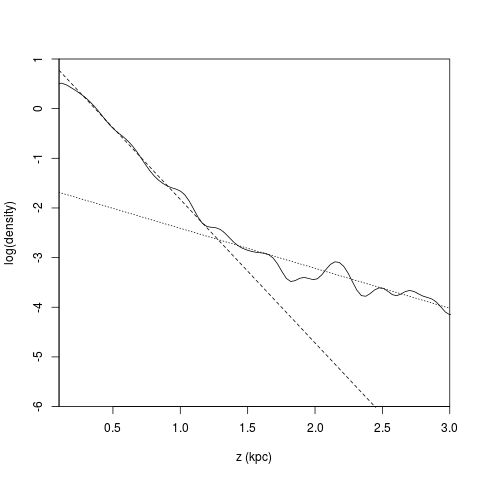}
 \caption{Density profile of the particles of the \cite{Fux99} model (disc + spheroid) in the solar cylinder. The fitted lines corresponds to exponential profiles with scale heights of 350~pc (dashed) and 1250~pc (dotted). } 
 \label{fig:FuxThickDisc}
 \end{figure}

There exist a large debate currently on whether the metal-poor component seen in the bulge is a classical bulge or the thick disc. The density profile of the spheroid component of \cite{Fux99} has been chosen in order to represent both the nuclear bulge and the stellar halo (looking in particular at RRLyrae and globular cluster density profiles). Indeed the outer bulge  has often been associated with the halo \cite[see e.g. the discussions in][]{IbataGilmore95}. 
Actually this spheroid component fits very well also the thick disc profile, as illustrated in Fig. \ref{fig:FuxThickDisc} which can be compared to the famous figure of \cite{GilmoreReid83}. In a solar neighbourhood Toomre diagram the spheroid particles lie within the thick disc and halo area. The discussions in \cite{GilmoreReid83} presents clearly this ambiguity between an exponential decrease perpendicular to the Galactic plane and the power law decrease from the Galactic Centre in a moderately oblate spheroid and already discuss a possible relation between the thick disk and the bulge. The dichotomy between classical bulge and thick disc interpretation for the metal-poor component is not obvious.
The importance of the actual density profile used for the inner regions is also illustrated in the study of \cite{Robin14} who found that a shorter scale length for the thick disc allows to nicely fill the place taken by the classical bulge population in their previous model \cite{Robin12}. \\

 Comparisons of our bulge to external galaxies also provides interesting insight, although one must take into account the very different observational biases between our detail star-by-star analysis and the global view of external galaxies. A classical bulge embedded within a peanut dominated profile is even more difficult to study in detail. Still the coexistence of classical and secularly evolved bulge has been observed in external galaxies \citep{Prugniel01,Peletier07,Erwin08}. Considering that the bulge is mainly old, observing high-redshift galaxies, e.g. before bar formation, is also very informative. See e.g. \cite{Kormendy15} for a recent review. 
 
\section{Conclusions\label{sec:conclu}}

The overall bulge follows the shape and kinematic signature of a secular bar, as predicted by dynamical models. Within this overall bulge, sub-divisions have been observed. 
In particular, adding metallicity information to the kinematics allows to highlight a metal-poor population presenting a kinematically distinct signature, being more centrally concentrated and extending further from the Galactic plane. This old population, consistent with a short time-scale formation through its $\alpha$-elements enrichment, should most likely have been present before the bar formation and with already an extended density profile and hot kinematics in order not to be as strongly influenced by the bar formation as the disc. However the exact density profile of this component and its links with the other Milky Way stellar populations are still largely unknown. 

Bulge density, sub-population proportions and kinematics change with sky direction and target distance.  A single population can also present internal gradients (see e.g. \citealt{Grieco12} discussing a metallicity gradient within a classical gravitational gas collapse component and  \citealt{MartinezValpuestaGerhard13} discussing a metallicity gradient in a secular bulge). Target selection bias (including biases induced by the extinction) complicates the detailed comparison with models. On the other side, the number of parameters at play within the models is very large. 

To understand the formation scenario of the bulge and its link with the populations observed locally, large homogeneous surveys providing both kinematics and chemistry are needed, covering not only the bulge but also the inner disc and the transition area with the thick disc and the inner halo. Gaia will soon provide excellent photometry and proper motions along the full bulge giant branch as well as distances for the brightest stars. Combined with other complementary photometric surveys (VISTA, LSST) and the new large spectroscopic surveys under-way (APOGEE, Gaia-ESO survey, HERMES) or planned (MOONS, 4MOST), the bulge formation history should be soon much tightly constrained.

\bibliographystyle{apj}
\bibliography{bulge_review.bib}

\end{document}